\documentclass[apj,twocolumn]{openjournal}

\usepackage{xcolor}
\usepackage{textgreek}
\usepackage[utf8]{inputenc}
\usepackage[english]{babel}

\usepackage{hyperref}
\hypersetup{
    unicode, 
    colorlinks=true,
    linkcolor=linkcolor,
    citecolor=linkcolor,
    filecolor=linkcolor,
    urlcolor=linkcolor,
}
\usepackage{color,colortbl}
\definecolor{linkcolor}{rgb}{0.0,0.3,0.5}
\usepackage{tensind}
\tensordelimiter{?}
\DeclareGraphicsExtensions{.bmp,.png,.jpg,.pdf}
\usepackage{verbatim}
\usepackage[normalem]{ulem}
\usepackage{orcidlink}
\usepackage{soul}

\graphicspath{ {./figs/} }

\usepackage{gensymb}

\newcommand{\mockalph}[1]{}

\begin{document}
\title{Understanding pulsar magnetospheres with the SKAO}

\author{L.~S. Oswald$^{1,2}$, A. Basu$^{3}$, M. Chakraborty$^{4}$, B.~C. Joshi$^{5,6}$, N. Lewandowska$^{7}$, K. Liu$^{8,9,10}$, M.~E. Lower$^{11}$, A.~Philippov$^{12}$, X. Song$^{13}$, P. Tarafdar$^{14}$, J. van~Leeuwen$^{13}$, A.~L. Watts$^{15}$, P. Weltevrede$^{3}$, G. Wright$^{3}$, J.~Ben\'a\v{c}ek$^{16}$, A. Beri$^{17,18,1}$, S. Cao$^{19}$, P. Esposito$^{20}$, F. Jankowski$^{21}$, J. C. Jiang$^{9,10}$, A. Karastergiou$^{2}$, K.~J.~Lee$^{19, 22}$, N. Rea$^{23,24}$, D. Vohl$^{13}$} 
\author{The SKA Pulsar Science Working Group}

\affiliation{$^{1}$ School of Physics \& Astronomy, University of Southampton, Southampton SO17 1BJ, UK}
\affiliation{$^{2}$Department of Astrophysics, University of Oxford, Denys Wilkinson Building, Keble Road, Oxford OX1 3RH, UK}
\affiliation{$^{3}$ Jodrell Bank Centre for Astrophysics, Department of Physics and Astronomy, The University of Manchester, Manchester M13 9PL, UK}
\affiliation{$^{4}$ Department of Astronomy, Astrophysics and Space Engineering (DAASE), Indian Institute of Technology Indore, Indore 453552, India}
\affiliation{$^{5}$ National Centre for Radio Astrophysics (TIFR), Pune 411007 India}
\affiliation{$^{6}$ Indian Institute of Technology, Roorkee 247667 India}

\affiliation{$^{7}$ Department of Physics and Astronomy, State University of New York at Oswego, Oswego, NY 13126, USA}
\affiliation{$^{8}$Shanghai Astronomical Observatory, Chinese Academy of Sciences, 80 Nandan Road, Shanghai 200030, China}
\affiliation{$^{9}$State Key Laboratory of Radio Astronomy and Technology, A20 Datun Road, Chaoyang District, Beijing, 100101, P. R. China}
\affiliation{$^{10}$Max-Planck-Institut f\"ur Radioastronomie, Auf dem H\"ugel 69, D-53121 Bonn, Germany}
\affiliation{$^{11}$ Centre for Astrophysics and Supercomputing, Swinburne University of Technology, PO Box 218, Hawthorn, VIC 3122, Australia}
\affiliation{$^{12}$ Department of Physics, University of Maryland, College Park, MD 20742, USA}
\affiliation{$^{13}$ ASTRON, the Netherlands Institute for Radio Astronomy, Oude Hoogeveensedijk 4,7991 PD Dwingeloo, The Netherlands}
\affiliation{$^{14}$ INAF - Osservatorio Astronomico di Cagliari, via della Scienza 5, 09047 Selargius (CA), Italy
}
\affiliation{$^{15}$ Anton Pannekoek Institute for Astronomy, University of Amsterdam, Science Park 904, 1098XH Amsterdam, the Netherlands}
\affiliation{$^{16}$ Institute for Physics and Astronomy, University of Potsdam, 14476 Potsdam, Germany}
\affiliation{$^{17}$ Indian Institute of Science Education and Research (IISER) Mohali, Punjab 140306, India}
\affiliation{$^{18}$ Indian Institute of Astrophysics, Koramangala II Block, Bangalore-560034, India}
\affiliation{$^{19}$ Department of Astronomy, School of Physics, Peking University, Beijing 100871, China}
\affiliation{$^{20}$ Scuola Universitaria Superiore IUSS Pavia, Palazzo del Broletto, Piazza della Vittoria 15, I-27100 Pavia, Italy}
\affiliation{$^{21}$ LPC2E, OSUC, Univ Orleans, CNRS, CNES, Observatoire de Paris, F-45071 Orleans, France}
\affiliation{$^{22}$ National astronomical observatories, Chinese Academy of Sciences, Beijing, China}
\affiliation{$^{23}$ Institute of Space Sciences (ICE-CSIC), Campus UAB, C/ de Can Magrans s/n, Cerdanyola del Vallès (Barcelona) 08193, Spain}
\affiliation{$^{24}$ Institut d'Estudis Espacials de Catalunya (IEEC), 08034 Barcelona, Spain}

\email{L.S.Oswald@soton.ac.uk}

\begin{abstract}

The SKA telescopes will bring unparalleled sensitivity across a broad radio band, a wide field of view across the Southern sky, and the capacity for sub-arraying, all of which make them the ideal instruments for studying the pulsar magnetosphere. This paper describes the advances that have been made in pulsar magnetosphere physics over the last decade, and details how these have been made possible through the advances of modern radio telescopes, particularly SKA precursors and pathfinders. It explains how the SKA telescopes would transform the field of pulsar magnetosphere physics through a combination of large-scale monitoring surveys and in-depth follow-up observations of unique sources and new discoveries. Finally, it describes how the specific observing opportunities available with the AA* and AA4 configurations will achieve the advances necessary to solve the problem of pulsar radio emission physics in the coming years.

\end{abstract}


\maketitle



\section*{Introduction}

Understanding the physics of the pulsar magnetosphere has been a key goal since the discovery of pulsars over half a century ago. Observations of pulsars in the radio, complemented by multiwavelength studies, reveal the relationships between the extreme magnetic and gravitational field strengths, electron--positron plasmas, quantum-electrodynamic particle processes, and particle acceleration mechanisms in the magnetosphere around the neutron star, and the connection between the pulsar magnetosphere and its interior. The study of pulsar magnetospheres is therefore a direct probe of the laws of physics at their extremes.

Radio observations of a pulsar reveal a wealth of information about the neutron star magnetosphere through a wide range of features, including the intensity and polarization of its pulses; the properties  of the integrated pulse profile formed by summing many pulses together; and the variability of these properties across radio frequency, from pulse to pulse, over long time periods, and across the pulsar population. Modern telescope advances, such as the construction of the MeerKAT \citep{Jonas2009} and FAST \citep{Nan2011} radio telescopes, have led to a rapidly growing pulsar population and the ability to collect high quality follow-up observations of this population, enabling advances in a statistical description of observed pulsar radio properties. The predicted elevation of our observational capacity through the SKA telescopes is expected to have a transformational effect on the remaining open questions about pulsar magnetosphere physics.

In this paper, we review the impressive advances in pulsar magnetosphere science over the last decade, focusing on observations and simulations of pulsar radio emission, and lay out the opportunities for continued scientific development in this field that will be achievable with the arrival of the SKA radio telescopes. We address these advances in the context of these five key science questions, identifying the open questions yet to be resolved and how the SKA can address these. The questions are as follows:
\begin{enumerate}
    \item What is the geometry of the neutron star magnetic field?
    \item What is the intrinsic emission spectrum of a pulsar?
    \item What are the origins of the time-variability of pulsar radio emission and spin-down?
    \item What is the global physics of the magnetosphere?
    \item How do pulsars evolve across the population and over their lifetimes?
\end{enumerate}

The paper ends with a discussion of the opportunities presented by the upcoming SKA telescopes for addressing these questions, and our predictions for the future of pulsar magnetosphere science. 

\section{What is the geometry of the neutron star magnetic field?} \label{sec:geom}

As a pulsar rotates, its radio beam sweeps through space, following a path defined by the angle of misalignment between the beam and the rotation axis. Every observable of a pulsar magnetosphere is defined by this geometrical set-up. Understanding a pulsar's geometry is therefore fundamental to resolving the bigger picture of how its magnetosphere is behaving.

\subsection{Constraining the inclination angle of the dipolar magnetic field} \label{sec:constrainalpha}

The linear polarization position angle (PA) of the pulsar radio beam has long been used to constrain the neutron star magnetic field geometry through the rotating vector model \citep[RVM, ][]{Radhakrishnan1969}, building on the assumptions that the neutron star magnetic field is dipolar, and that the PA carries the imprint of the magnetic field line structure at the point in the magnetosphere where the radio emission is produced. Constraining the field geometry through the RVM is however difficult in most cases, due to the small duty cycles of many pulsars leading to covariance of the model parameters \citep{Rookyard2015}, and the presence of more complex PA profiles than can be fit by the model \citep{Oswald2023a}. The RVM is better constrained for pulsars with interpulses \citep{Johnston2019}, and modern surveys have enabled a statistical picture of RVM fitting for the population as a whole, even if individual measurements are poorly constrained \citep{Johnston2023}. Although 59 per cent of pulse profiles are not successfully fit by the RVM due to additional PA complexity \citep{Johnston2023}, \cite{Johnston2024} found that this failure rate is reduced to 5 per cent if an integrated pulse profile is formed only of bins with linear polarization fraction greater than 80 per cent, following the prescription developed by \cite{Mitra2023}. This points towards considerable future advancements in understanding neutron star magnetic field geometries on a population scale, as well as revealing insights into other magnetospheric propagation effects discussed further in Section \ref{sec:morphpolspec}. Studying magnetic field geometries of the population as a whole will help us address the question of whether the magnetic axis aligns with the rotational axis over time, as is hinted at by studies of pulsars with interpulses \citep{Weltevrede2008}.

Even though the neutron star magnetic field is a 3-dimensional object -- arguably even  4-dimensional if one includes the magnetospheric changes with time \citep[e.g.,][]{lhk+10} -- our line of sight generally only provides information in a lower-dimensional slice, giving a one-dimensional cut through the radio beam as the pulsar rotates. In some pulsars that are part of a binary system, however, the general relativistic effect of geodetic precession allows for a more complete study of a pulsar's emission beam. One such source is PSR\,J1906+0746 \citep{vanLeeuwen2015}.  Over the course of several years, the changing line of sight allows us to create emission beam maps of both poles of this pulsar \citep{Desvignes2019}, including a verification of the RVM model. Pulsar surveys with SKA will find a number of new, similar systems. The different geometries of those will help complete our understanding of the pulsar magnetic field geometry, and test the extent to which the dipolar field assumption holds across the pulsar population. 

\begin{figure}
    \centering
    \includegraphics[scale=0.5]{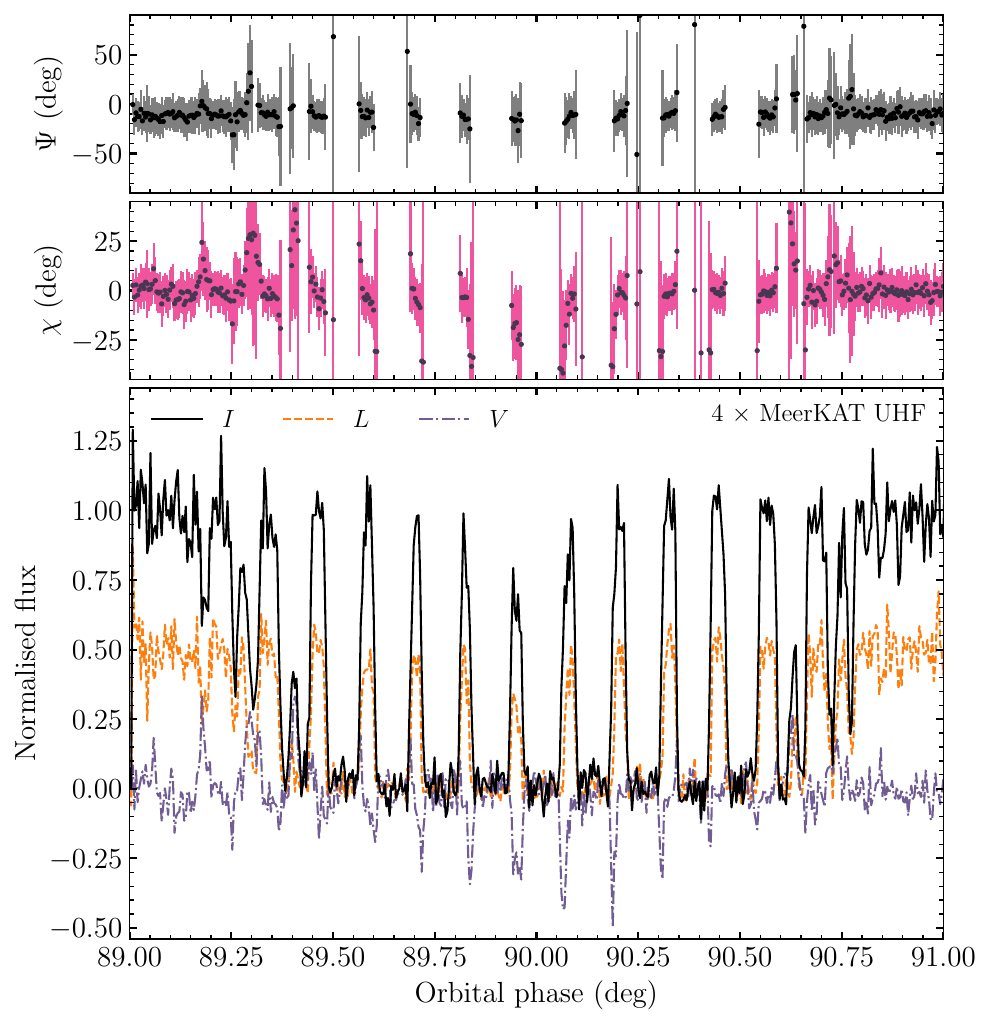}
    \caption{Average of four MeerKAT UHF polarization light curves of PSR~J0737$-$3039A being eclipsed by the truncated dipolar magnetosphere of PSR~J0737$-$3039B. Top panel shows the linear polarization position angle ($\Psi$), middle is the ellipticity angle ($\chi$) and the lower panel depicts the changes in total intensity ($I$; black solid line), total linear polarization ($L$; dashed orange line) and circular polarization ($V$; dash-dotted purple line).}
    \label{fig:double_psr_eclipse}
\end{figure}

In addition to being an excellent testing ground for fundamental physics, the double pulsar system PSR~J0737$-$3039A/B provides a unique direct view into the magnetosphere of an active radio pulsar. The orbit of the system is viewed almost perfectly edge-on, resulting in 30-40\,s duration eclipses at superior conjunction, where radio emission from the fast-spinning `pulsar A' is obscured by the opaque plasma surrounding the slow-spinning `pulsar B' \citep{lbk+2004}. As shown in Figure~\ref{fig:double_psr_eclipse}, the radio light-curves of the eclipses display periodic peaks and troughs throughout the eclipse envelope that are modulated at both once and half the 2.76\,s spin period of pulsar B \citep{mll+2004}. These modulations are result from synchrotron absorption by the relativistic pair-plasma trapped within the truncated dipolar magnetic field of pulsar B \citep{lt05}. Secular changes in the modulation pattern detected by the Green Bank Telescope were linked to geodetic precession of the pulsar spin-axis about the binary total angular momentum vector, enabling a sixth test of general relativity \citep{bkk+08}. Recent observations with MeerKAT improved upon these measurements, with advances in statistical inference techniques allowing for several covariances between the orientation of the system and viewing geometry to be overcome \citep{lks+24}.

\subsection{Non-dipolar field geometries and multi-wavelength observations} \label{sec:nondipolar}

The ultraviolet (UV) component of pulsar electromagnetic spectrum can originate from thermal radiation of hot spots located on the star surface, or by other types of non-thermal radiation from the magnetosphere, or a combination of both \citep{Iniguez-Pascual2022,Takata2017}.
However, the relation between the position of UV emission regions and the magnetospheric plasma and field structures remains uncertain.
By combining radio and UV observations, a more refined understanding of emissions and the emission mechanisms will be facilitated \citep{Krticka2024}.
The primary open question addressed by the outcomes is the long-standing issue between the dipolar or multipolar shape of the magnetic field, and how the shape impacts the magnetospheric processes \citep{Yao2018}.

Prior to the launch of the Fermi Gamma-ray Space Telescope in 2008, fewer than a dozen gamma ray pulsars had been discovered. With three iterations of the Fermi LAT Catalog of Gamma-ray Pulsars published since then, that number is now approximately 340, almost 10 per cent of the known radio pulsar population \citep{Smith2023}. Gamma-ray pulse profiles are usually broad and double-peaked, and tend to be misaligned in time with respect to the radio pulse profile \citep[e.g.][]{Smith2023}. It is likely that, unlike radio emission which is produced in the polar cap region, gamma-rays from pulsars are largely generated in the current sheet beyond the pulsar light cylinder \citep{Cerutti2024}. Gamma-ray emission from pulsars has been detected into the TeV range, with the highest energy pulsar emission detected being a 20~TeV component from the Vela pulsar \citep{H.E.S.S.Collaboration2023}, information which is helping to constrain our understanding of the particle acceleration mechanisms required to produce such high energy emission.

PIC simulations have shown that the percentage of the Poynting flux converted into pairs/gamma rays in the current sheet varies with magnetic field geometry: for an inclination angle of 90$\degree$ it is roughly 1 to 2 per cent, whereas for zero inclination angle the amount is closer to 10 per cent \citep{Cerutti2020}. This means that measuring the magnetic field inclination for gamma-ray pulsars would be a particularly powerful tool for understanding the pulsar magnetosphere \citep{Hakobyan2023}. It is also possible that, under certain theoretical assumptions, gamma-ray observations of pulsars might provide an additional constraint on pulsar magnetic field geometries, when considering these measurements alongside fits to the radio polarization \citep{Petri2021}. 

A number of rotation-powered radio pulsars are also X-ray pulsars, as magnetospheric currents heat the stellar surface at the magnetic poles to temperatures where they emit thermal radiation in the X-ray band \citep{Harding01}. Radiation from the hot poles is modified by special and general relativistic effects as it propagates out of the star's gravitational potential well towards the observer; by modelling the resulting X-ray pulse profile, one can infer the mass and radius of the pulsars.  This in turn can yield information about the dense matter equation of state. However the technique can also let us infer the properties of the hot poles (their shape, size, temperature distribution and location), effectively mapping the footprints of the magnetic field. 

Large X-ray spectral timing data sets from NASA's Neutron Star Interior Composition Explorer \citep[NICER][]{Gendreau16} are now enabling pulse profile modelling for a number of bright rotation-powered X-ray pulsars \citep[see][and references therein for more detailed description of the technique]{Bogdanov19b,Bogdanov21}.  The fact that these sources are also radio pulsars is critical; in addition to providing the ephemeris to generate the phase-resolved X-ray pulse profile, radio timing (if the source is in a binary) can also provide priors on mass and inclination, all key parameters in the modelling. To date pulse profile modelling results using NICER data have been reported for five sources:  the  2.1 $M_\odot$ pulsar PSR J0740+6620 \citep{Cromartie20,Fonseca21,Riley21,Miller21,Salmi22,Salmi23,Salmi24a,Dittmann24,Hoogkamer25}; the 1.4 $M_\odot$ pulsar PSR J0437-4715 \citep{Reardon24,Choudhury24}; the isolated pulsar PSR J0030+0451 \citep{Bogdanov19a,Riley19,Miller19,Vinciguerra24}; the binary pulsar PSR J1231-1411 \citep{Salmi24b,Qi25} (for which the mass is poorly constrained); and the 1.4 solar mass pulsar PSR J0614-3329 \citep{Miles2025, Mauviard2025}.

For all of these sources, the inferred magnetic pole properties point to a complex field geometry that is not consistent with a centered dipole. While two hot magnetic poles appear to be sufficient to describe the data, they may have complex shapes (such as long elongated arcs) or temperature distributions, and are often far from antipodal. An off-centered dipole or quadrupolar component to the surface field appears to be required to explain the data \citep[see e.g.][]{Gralla17,Lockhart19,Bilous19,Chen20,Kalapotharakos21}. This poses new challenges for our understanding of the pulsar emission mechanism (since this same magnetosphere must be able to explain the radio and gamma-ray emission), and neutron star evolution.  One open question is whether these complicated field geometries formed at birth, or whether they developed during the accretion and recycling process.

\subsection{Summary}
In summary, multi-wavelength observations of pulsars reveal the complexities of pulsar magnetic field geometries, and how they evolve across the pulsar population and over a pulsar's lifetime. Key questions yet to be answered unambiguously are as follows: how far can the dipolar magnetic field assumption be extended, and at what point do non-dipolar field components become important? Do pulsars align their radio beams with the rotation axis over time? How does the geometry impact what is observable at different wavelengths, and what can we learn about the three-dimensional structure of the radio beam? Understanding the magnetic field geometry is the fundamental building block on which the rest of our picture of the pulsar magnetosphere structure and behaviour is constructed.

\section{What is the intrinsic emission spectrum of a pulsar?}
The intrinsic emission spectrum from a source is a key diagnostic of the astrophysical process, or processes, driving the emission. Combining this with the intrinsic luminosity gives the combination of the energetics of the source and how that energy budget is being converted to photons. For pulsars, our constrained view of the beam structure along the line of sight, plus the geometry of the magnetosphere, both affect what spectrum we observe. These factors also limit our ability to estimate the true luminosity of the source in the radio, because of the difficulty of trying to extrapolate from the radio emission observed along the line of sight to the total radio emission produced by the whole beam. Our understanding of the intrinsic spectrum is further complicated by the impact of multiple processes in the magnetosphere which alter the radio emission ultimately observed. Broad-band observations of pulsar radio emission are therefore very important for disentangling these processes and understanding the physics at play.

\subsection{Limitations of simple pulsar models in explaining broad-band observations}

Simple descriptions of the standard radio pulse profile, and the associated empirical models of generation and propagation in the pulsar magnetosphere, are increasingly being challenged by the sheer variety seen in modern observations of the pulsar population. Common assumptions behind such models: that pulse profiles are often symmetric, tend to widen with decreasing frequency, exhibit high linear polarization and negligible circular polarization, and have an S-shaped PA, are rarely all simultaneously true for a given source. This presents challenges to models of the emission beam structure as consisting of either hollow cones, patches or fan beams; of the presence or not of radius-to-frequency mapping; and on our ability to infer the height above the neutron star surface at which radio emission is produced by measurements of aberration and retardation effects. 

\begin{figure}
    \centering
    \includegraphics[width=\columnwidth]{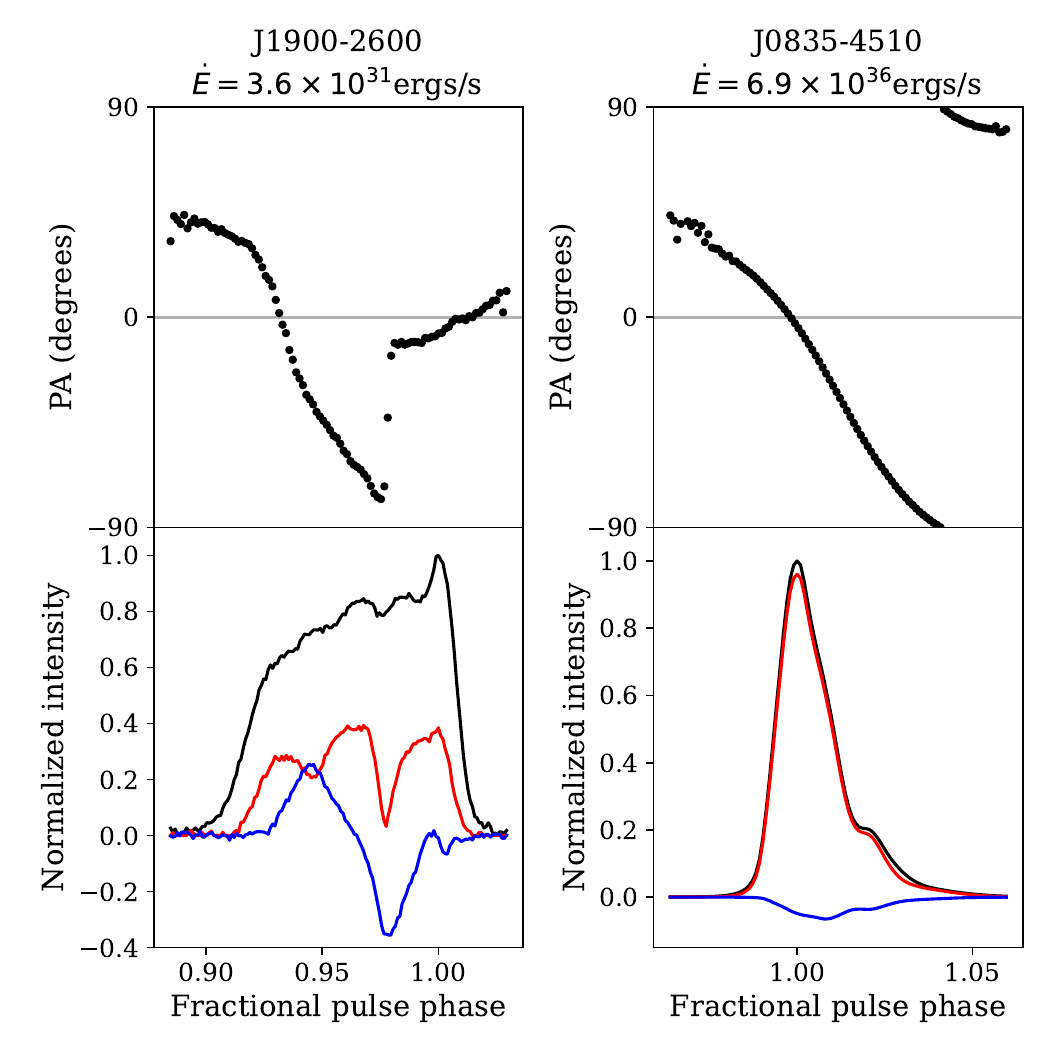}
    \caption{Two examples of integrated pulse profiles of pulsars observed with the Murriyang radio telescope using its Ultra-Wideband receiver. The full frequency resolution of the receiver (704--4032 MHz) is split into 8 subbands, and the pulse profiles displayed here are taken from the subband centred at 1400~MHz. In each case the top subplot shows the position angle (PA) of the linear polarization, and the bottom subplot shows the total intensity (black), linear polarization (red) and circular polarization (blue) of the pulse profile.}
    \label{fig:EdotPulsars}
\end{figure}

Integrated pulse profiles of pulsars display a wide range of pulse profile shapes (including single/double/multiple/blended component profiles) \citep{Vohl2024}, pulse widths \citep{Posselt2021}, and polarization fractions that average 20--30 per cent linear polarization and 5--10 per cent circular polarization \citep{Gould1998, Oswald2023a} but can range from near-100 per cent linear polarization to completely depolarized \citep{Oswald2023a}. Examples of two integrated polarized pulse profiles with very different profile shapes and polarization properties are shown in Fig. \ref{fig:EdotPulsars}. These examples also have very different values of spin-down energy $\dot{E}$, and it has been shown more widely that all of these features show evolving trends across the population, particularly with respect to $\dot{E}$ \citep{Oswald2023a}.

Advances in wide-band telescope technology such as the Ultra-Wideband receivers on Murriyang \citep{Hobbs2020}, Effelsberg \citep{Dunning2022} and GBT \citep{Lynch2023}, complemented by the extension of observations to lower frequencies with the u-GMRT \citep{Gupta2017}, LOFAR \citep{VanHaarlem2013} and NenuFAR \citep{Boissier2012, Bondonneau2021} telescopes, have enabled a broad-band picture of pulsar radio emission. Pulsar flux spectral index distributions have been measured to average $-1.6\pm0.3$ \citep{Jankowski2018} and $-1.8\pm0.8$ \citep{Posselt2023} for surveys of normal pulsars with the Murriyang and MeerKAT radio telescopes respectively. 

\subsection{Spectral behaviour of pulsar morphology and polarization} \label{sec:morphpolspec}

Polarized pulse profiles are found to display complex frequency evolution of the PA and circular polarization fraction for many pulsars, with a tendency towards overall depolarization at higher frequency \citep{Oswald2023a}. Although many pulsars display widening of their pulse profiles with decreasing frequency below 1~GHz, a feature commonly explained by the concept of ``radius-to-frequency mapping'' \citep[RFM;][]{cor78}, several more show either no frequency evolution, or the inverted effect, with widening at higher frequencies \citep{Posselt2021}.

Shared similarities in pulse profile shape — and in their evolution across frequencies — are observed for pulsars spanning a wide range of periods and period derivatives, from millisecond to normal pulsars. A multi-frequency minimum spanning tree analysis demonstrated that pulsars can be sequenced by morphological similarity \citep{Vohl2024}. Moreover, akin to the period—width relation \citep[e.g.][]{Johnston2019, Karastergiou2024}, a sub-sample of 90 pulsars from the European Pulsar Network database sequenced by shape hints at a mild correlation with the pulsar period. On a related note, it has been shown that for energetic pulsars, linear polarization is known to correlate strongly with spin-down energy loss rate \citep[$\dot{E}$,][]{2008MNRAS.391.1210W}, whereas profile morphology has a more complex relationship with $\dot{E}$. Although it was found that pulsars with lower $\dot{E}$ tend to have more complex profiles \citep{Oswald2023a}, the relation between profile morphology and $\dot{E}$ appears to evolve with heteroscedasticity along this morphological sequence (i.e. the scatter in $\dot{E}$ increases along the sequence) \citep{Vohl2024}. SKA observations will help disentangle morphology and polarization dependencies on key physical parameters.

The presence of jumps of 90$\degree$ in some PA profiles has led to the hypothesis of birefringence in the pulsar magnetosphere, causing the production and independent propagation of two orthogonally polarized modes of emission \citep[OPMs, e.g. ][]{Barnard1986}. More complex pulse profile behaviour includes having multiple blended components in the pulse profile; the presence of both orthogonal and non-orthogonal jumps and modulations in the PA profile; exhibiting complex frequency dependence of the polarization properties; and possessing a significant polarization fraction which is circular. All of these features tend to occur together \citep{Oswald2023a}, particularly at higher radio frequencies and for pulsars with lower spin-down energies, and can be modelled empirically through coherent or partially-coherent combination of orthogonal polarization modes \citep{Dyks2019, Dyks2021, Oswald2023b}. This indicates that these features originate as propagation effects in the pulsar magnetosphere \citep[e.g.,][]{Arons1986,Petrova2000,Wang2010,BeskinPhilippov2012}, which itself displays evolving properties over a neutron star's lifetime. 

\subsection{Intrinsic luminosities of radio pulsars}

To fully understand pulsar magnetosphere physics we need to know the full energy budget available (which will arise from the spin-down energy and some additional contribution from the magnetic field), how it transforms into photons, and what the photon emission mechanism or mechanisms are in each waveband. This means that it is important to understand the efficiency of radio pulsars, namely what fraction of spin-down energy is converted to radio luminosity and to what extent this is consistent across the pulsar population. Understanding the intrinsic radio luminosity of pulsars is also a highly important constraint on predicting how many pulsars will be discovered with the SKA telescopes, not only in terms of identifying the distances to which we are sensitive to faint sources, but also in terms of constraining our understanding of the beaming of the pulsar radio emission and how this impacts what fraction of sources will point towards the Earth. Our ability to understand both the radio luminosity and the total luminosity of pulsars is limited by geometry: not only due to the complexity of radio beam structure, but also in terms of relating how the geometry affects which emission we do and do not see across all wavebands (for example in cases where the radio emission is observable but the gamma ray emission points away from the observer, or vice versa). Understanding both the magnetic field geometry, as discussed in Section \ref{sec:geom}, and the radio beam structure, which is revealed by broad-band observations as described above, are therefore vital to building up a full picture of the intrinsic pulsar luminosity in the radio. In turn, it is then important to use this information to understand how luminosity varies between different populations of neutron stars. \cite{Szary2014} found that pulsar radio emission efficiency is inversely linearly correlated with spin-down energy and linearly correlated with characteristic age, implying that there is a weak, if any, correlation between luminosity and pulse period and spin-down. Comparing millisecond pulsars to ordinary pulsars, \cite{Kramer1998} suggested that millisecond pulsars were less luminous than ordinary pulsars, and suggested that their emission beams are narrower. However, a recent study by \cite{Karastergiou2024} pointed out that using pseudo-luminosity measurements to compare millisecond and ordinary pulsar populations is not appropriate for making inferences about intrinsic luminosity, and that in fact on scaling the pseudo-luminosity to account properly for the pulse duty cycle and beam solid angle the inferred intrinsic luminosities of the two populations are indistinguishable. Such studies take us closer to a global understanding of pulsar radio energetics for the whole population. 

\subsection{Pulsar radio emission at higher frequencies}

To date, higher frequency ranges, in particular those above 10\,GHz, have barely been explored. Under a RFM scenario, we might expect higher frequency radiation to be generated at a lower emission height, meaning that higher frequency observations could be used to test this picture more thoroughly. High-frequency spectral windows could therefore be very insightful in probing deep into the polar cap region and its emission beam pattern. It can also be used to search for emission components that are prominent only in the high-frequency bands \citep{hej16,lyw+19}. This may be used to test models that predict a different emission mechanism corresponding to these high-frequency emission components \citep[e.g.,][]{lyu13,ckl01}. The SKA Band 5b will for the first time, offer the requested sensitivity to enable high-radio-frequency ($>10$\,GHz) study of pulsar emission for a large number of samples. 

\subsection{The frequency-dependent impact of the Interstellar Medium}

Understanding the frequency-dependence of pulsar radio emission requires a simultaneous understanding of the frequency-dependent impacts of the interstellar medium (ISM). The most common observational impacts of the propagation of pulsar radio emission through the ISM are those of dispersion and Faraday rotation, with many pulsars exhibiting strong interstellar scattering and scintillation, and some sources being further impacted by the Solar wind. The need for careful constraint of these effects becomes ever more important with the advent of Ultra-Wideband receivers on radio telescopes enabling broad-band observations \citep{Oswald2020}. The frequency-dependent profile morphology coupled with variable scatter-broadening impacts the precision of dispersive effects thereby introducing a variable noise source in the pulse timing delay \citep{shml2021,sjk+2024}. Therefore, a better understanding of these effects is important for a sensitive detection of gravitational wave background through pulsar timing array experiments \citep[see][from this special issue]{Shannon2025_SKA_SKAPTA}. It simultaneously presents a further scientific opportunity to use pulsars to probe galactic plasmas on a wide range of scales \citep[see][from this special issue]{Tiburzi2025_SKA_Plasma, Xu2025_SKA_GMF}.

\subsection{Summary}
In summary, there are multiple physical mechanisms which impact the pulsar radio emission properties ultimately observed, and broad-band observations of pulsars enable us to make progress in disentangling the impacts of these different physical mechanisms. The large number of pulsars that will form the SKA sample and its data quality, particularly full-Stokes observations observed with the high instantaneous band-widths of both SKA-Mid and SKA-Low, will be key to further the study of such relationships. This will allow detailed mapping of integrated profile features onto trajectories on the Poincar\'e polarization sphere and enable statistical studies to investigate emission mechanisms and propagation through the magnetosphere and interstellar medium.

\section{What are the origins of the time-variability of pulsar radio emission and spin-down?} \label{sec:timevar}

An important puzzle presented to us is why pulsars are variable at all timescales imaginable. What are the physical conditions and mechanisms that drive the variability? Here we discuss the observational impact of time-variability of pulsar radio emission on both long and short timescales. We also discuss its origins in relation to both profile changes and spin-down, and the correlation between the two.

\subsection{Single pulse variability} \label{sec:subpulse}

Individual pulses from pulsars show complex substructures - termed subpulses - which appear to march in phase within the pulse window, a phenomenon known as subpulse drifting. The subpulse drifting in pulsars shows a range of evolution, primarily displaying a stable drifting behaviour \citep{Weltevrede2007, Song2023}. The phenomenon has been described using a rotating carousel model of a discrete number of `sparks’ (electrical discharges) just above the neutron star surface near the magnetic poles \citep{rs75}. However, large-scale observations of the pulsar population have unveiled a wide range of pulsars displaying atypical subpulse drifting characteristics, which cannot be fully explained using the standard scenario. For example, for a subset of pulsars, the phenomenon of bi-drifting was observed, which was attributed to the physics of inner acceleration gap \citep{qlz+04} or non-circular spark motions \citep{wright2017}. \citet{szary2022} reported the detection of sudden drift rate reversal which could be explained by a modified carousel model where the sparks rotate around the location of the electric potential extremum of the polar cap. Another remarkable feature observed from a subset of the subpulse drifting pulsars is rapid drift mode changing along with the steady evolution of the drift rate within a particular drift mode  \citep{mcsweeney2022, janagal2023}. A range of physical models involving variable drift rate and variable spark number/configuration have been evoked to explain such atypical properties. However, it should be noted that simulations of magnetospheres to date (discussed further in Section \ref{sec:globalstructure}) have not generated spark structures, meaning that the underlying mechanism driving subpulse drifting remains unexplained.

Concurrent observations of subpulse drifting over multiple frequencies can be useful to disentangle the beam geometry from the magnetospheric processes, as drift rate of subpulses provides a probe of the polar cap physics. These, as well as  non-simultaneous observations, suggest a broad-band nature of not only subpulse drift, but pulse nulling as well \citep{Weltevrede2006,Weltevrede2007,gjk+14,njmk17}. The wide-frequency coverage and sensitivity of SKA provides a unique opportunity to investigate the relationship of concurrent drift rate changes with profile modes and single pulse polarization to provide unique insight in the magnetospheric processes and polar gap physics.

Although subpulse drifting is widely reported in the literature, with at least 35 per cent of pulsars exhibiting observable drifting \citep{Song2023}, due to sensitivity constraints, full polarimetric investigations of subpulse drifting are limited. For a handful of pulsars, studies have revealed OPM switching in synchronization with the drifting sub-pulse modulation \citep{taylor1971, ramachandran2002, rankin2003, rankin2006, Edwards2004, Ilie2020}. \citet{primak2022} reported that for specific pulse longitudes, the polarization state of pulsar B1919$+$21 evolves synchronously with the pulsar drift period. The polarization view of the subpulse drifting modes in multiple frequency bands can be immensely valuable for addressing long-existing questions related to the subbeam configuration, evolution, and emission height dependence. The enhanced sensitivity and polarimetric capabilities of SKA hold immense potential for addressing these long-standing questions related to the physics of the pulsar magnetosphere. Furthermore, the subarray configuration capabilities of SKA can offer critical insight into the emission height dependence of the different spark configurations and their variation.

The individual pulses from pulsars show a considerable variation in their intensity and pulse width over a wide range of timescales, from nanosecond subpulse structure to modulation from pulse to pulse.
This modulation includes both short-term and long-term mode-changing, which has been found to show quasi-periodicity in pulsar J1326$-$6700 \citep{Wen2020}, and nulling, which may be an extreme form of mode-changing \citep[e.g.][]{njmk17,njmk2018}. 
\citet{bs2012} investigated the single pulse energy distribution and associated modulation index characteristics of 315 pulsars, and showed that about 40 per cent of pulsars behave in a distinct manner (log-normal pulse energy distribution) while the others show Gaussian or more complex behaviour. Extending such studies to include investigation of the spectral variation of the single pulses can offer vital clues about the instantaneous magnetospheric plasma activity \citep{Bilous2022,janagal2023a}.

Giant pulses are single pulses that show very high flux densities and are apparently non-periodic \citep{staelin}, follow a power-law in their amplitude distributions \citep{lundgren_1995}, show a phase-bound occurrence \citep{heiles_1970}, pulse widths in the range of a few microseconds to nanoseconds \citep{hankins_2003,soglasnov_2004} and follow a Poisson distribution \citep{lundgren_1995}. Originally discovered in the Crab pulsar \citep{staelin} such anomalous single pulses were also observed from recycled pulsars like PSR B1937+21 \citep{sallmen_backer_1995}. Since then a small group of pulsars has been reported to show radio giant pulses \citep{knight_2006} including band-limited giant pulses that bear a resemblance to some Fast Radio Bursts \citep{Geyer2021}. The mechanism driving giant pulses may not be the same as for ordinary single pulses, as discussed in Section \ref{sec:globalstructure}.

The shortest timescales of substructure within radio pulses from pulsars have commonly been termed micropulses or microstructure. They are of particular interest when considering the size scales driving pulsar radio emission, since a micropulse being produced on a very short timescale must correspond either to a small length-scale in the region where the radio emission is produced, or to highly relativistic beaming, constraining the radio emission mechanism \citep{Kramer2024}. In recent years, it has been found that microstructure is observed not only in ordinary pulsars, but also in all classes of radio-emitting neutron stars, from millisecond pulsars \citep{Liu2022} to magnetars \citep{Kramer2024}. These micropulses commonly appear quasi-periodically, and it has been found that the quasi-periodicity scales with the rotational period of the pulsar for a selection of sources with periods spanning more than six orders of magnitude \citep{Kramer2024}. This indicates the potential for a fundamental connection in the emission mechanism of radio pulsars across all of the sub-populations: canonical pulsars, millisecond pulsars, rotating radio transients (RRATs) and magnetars, with a further potential for this connection to be extended to Fast Radio Bursts, which is discussed further in Section \ref{sec:frb}. With the possible indication of such a universal connection, it has been proposed to rename microstructure to quasi-periodic sub-structure, to highlight the connection between quasi-periodicity and the rotation period. Studying quasi-periodic sub-structure for an increased number of sources must be done to test whether this connection holds for the pulsar population as a whole. This requires not only high time resolution but also high telescope sensitivity to be able to resolve the sub-structure. The SKAO telescopes will be ideal instruments to undertake such studies.

\subsection{Spin-down variability and profile changes}

Another source of pulsar time variability is glitches, whereby a pulsar's spin-down is interrupted by a sudden spin-up event, before relaxing back towards its original state \citep{Antonopoulou2022}. This is thought to originate from pinning and un-pinning of vortices in the superfluid neutron star interior, leading to exchanges of angular momentum between the interior and the crust. However, it has been shown that the impact of glitching is not restricted to the interior neutron star configuration, but also to the magnetosphere: \cite{Palfreyman2018} showed that, during and just after a glitch, the single pulses of the Vela pulsar showed nulling, pulse shape change, a drop in linear polarization, and late arrival in the pulse window. Studying the impact of glitching on single pulse behaviour requires high quality and high time resolution observations and so this is the only such case recorded to date, but monitoring with SKA telescopes would be expected to produce additional such cases for further study. As glitches are rare events, the wide sky coverage of SKA-low will be particularly useful for catching such events and for commensal high cadence monitoring.

At the other extreme of time variability, pulse profiles of at least some pulsars are known to be variable at decade-long timescales. Radio emission variability is known to be driven by powerful magnetospheric processes, as demonstrated by the discovery that radio emission mode changes of PSR B0943+10 are accompanied by X-ray state changes \citep{hhk+2013}. Furthermore, these processes are so powerful that radio emission variability is accompanied by spin-down rate changes of the neutron star, which implies that pulsar timing experiments are affected by these magnetospheric processes. These associated spin-down rate changes were first identified for pulsars that switch off their radio emission completely \citep{klo+06} and later for pulsars that have more subtle emission changes \citep{lhk+10}. Discrete sudden profile change events, followed by a long term recovery towards a normal profile, recently reported for PSR J1713+0747 \citep{ssj+2021,Jennings2024}, pose challenges for precision timing of such millisecond pulsars. An additional challenge is posed by observation duration dependent intrinsic variability of the pulsar signal manifesting as  stochastic wide-band impulse modulated self-noise or ``pulse jitter'' in pulsar timing \citep{ovsh+2011,lkl+2012,pbs+2021,khb+2024}. This emission variability is thought to be linked to changes in electron-positron pair creation in the magnetosphere, and thereby affecting the torque caused by the pulsar wind. 

\begin{figure}
    \centering
    \includegraphics[scale=0.318]{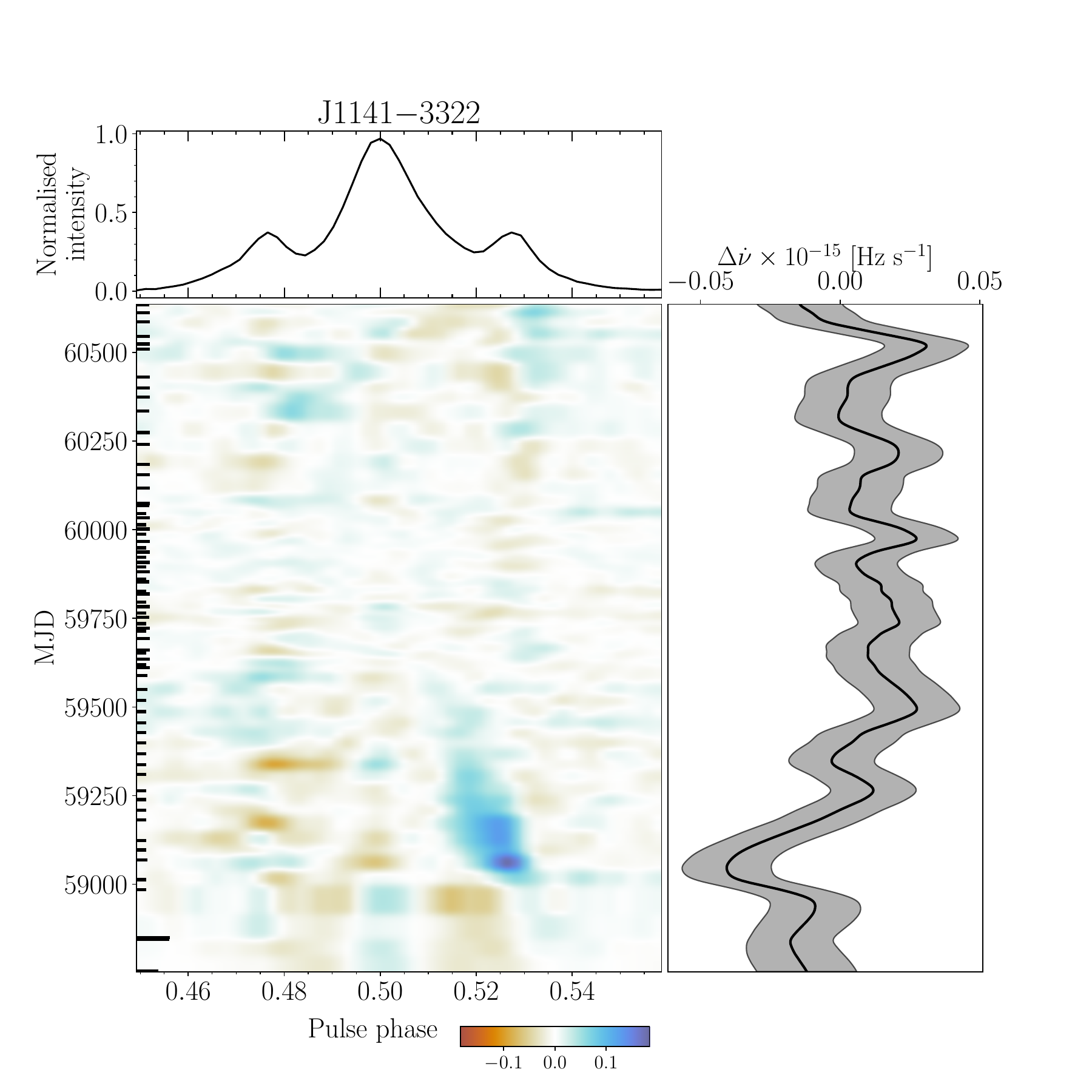}
    \caption{An example of a pulsar with profile variability as identified in the Thousand-Pulsar-Array project with MeerKAT. The figure is similar to that in \cite{bwk+24}, but includes more recent data. The top panel shows the average pulse profile of PSR J1141$-$3322. The colour map in the lower left panel shows the phase resolved temporal evolution of profile shape of the pulsar (for details see \citealt{bwk+24}). The lower right panel shows the variations of $\dot \nu$ over the secular spin-down rate. An excess of emission associated with the trailing component can be seen to be mildly correlated with the increase in spin-down rate.}
    \label{fig:long-term-prof-var}
\end{figure}

A sensitive and regular radio monitor campaign is key to identify the origins of pulsar radio emission variability on these long timescales. Indeed more recent campaigns with MeerKAT \citep{bwk+24} and Murriyang \citep{lkj+25} show that emission variability is common throughout the pulsar population. An example of such profile variability observed using the MeerKAT radio telescope in PSR J1141$-$3322 is shown in Fig.~\ref{fig:long-term-prof-var}. Identifying these systems will shed light on how radio emission, magnetospheric physics and pulsar timing are linked. The SKA will provide unprecedented pulsar timing sensitivity. Therefore understanding of emission variability will be crucial for the high precision timing experiments required to constrain GR and characterize the gravitational wave background.

\subsection{Summary}
As the number and quality of radio pulsar single-pulse observations increase, so too does the prevalence of time-variability observed in pulsar radio emission. It is increasingly clear that both subtle and dramatic time-variability is ubiquitous on all time-scales across the pulsar population. It is highly important that we understand the origins and behaviours of pulsar time-variability, not only to constrain the physics of the pulsar magnetosphere, but also to mitigate the impact of sudden time-variable events on millisecond pulsar timing \citep{Nathan2023,Jennings2024} and so advance the search for gravitational waves with radio pulsars. 

\section{What is the global physics of the magnetosphere?} \label{sec:globalstructure}

The majority of this paper has so far focused on the advances in understanding of the pulsar magnetosphere revealed by radio observations of pulsars. However, the radio emission is not generated in isolation: it is important to consider the wider context of the magnetosphere as a macroscopic structure. Multi-wavelength observations of pulsars in the UV, gamma and X-ray bands indicate more complex elements of pulsar magnetosphere physics are present, particularly in relation to potential non-dipolar field geometries  (see Section \ref{sec:nondipolar}). The best progress in obtaining a global picture of the pulsar magnetosphere is to pair observational progress with advances in simulating and modelling the pulsar magnetosphere and how it generates photons.

\begin{figure*}
    \centering
    \includegraphics[width=0.9\linewidth]{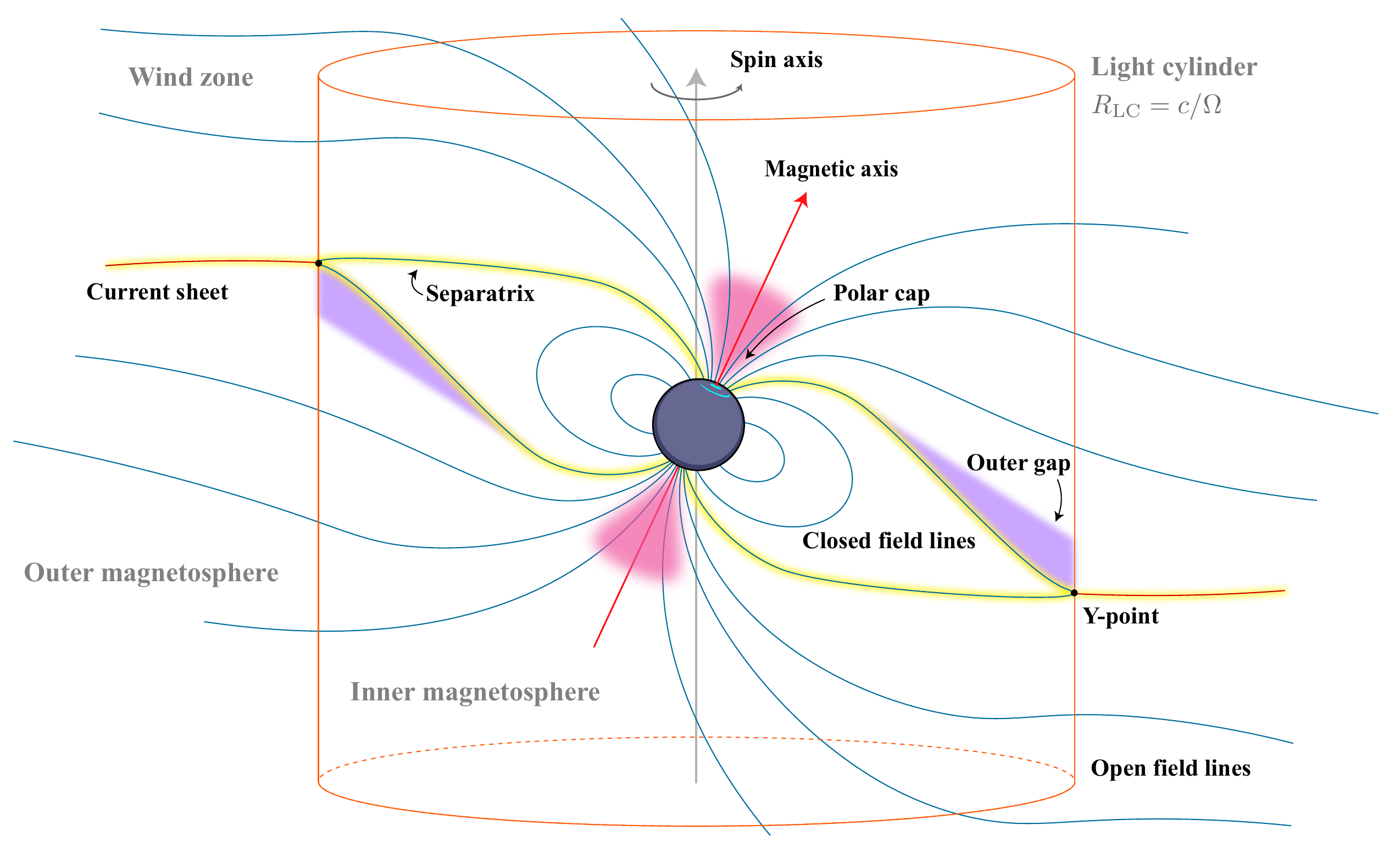}
    \caption{Illustration of the dipolar magnetic field of a pulsar where key regions of interest are highlighted. The inner and outer magnetosphere are defined by the regions within and beyond the light-cylinder radius, indicated by the orange cylinder, where the co-rotation velocity exceeds the vacuum speed of light. Electromagnetic radiation is generated in the shaded regions, with radio emission originating from the polar cap (pink) and potentially the separatrix (yellow). High-energy non-thermal emission, including gamma-rays, is thought to originate from the current sheet beyond the Y-point (yellow; as opposed to older models prescribing radiation from the ``outer gap'', purple). Thermal X-ray emission arises from the hotspots (light blue regions on the central neutron star) where energetic particles, which are produced by the pair production discharges, bombard the stellar surface in the open field line region.}
    \label{fig:cartoon}
\end{figure*}

\subsection{The canonical pulsar magnetosphere and the modern perspective from simulations}

The canonical picture of the pulsar magnetosphere is derived primarily from the Goldreich-Julian and Ruderman-Sutherland models \citep{gj69, rs75}. For a perfectly conducting magnetised sphere rotating in vacuum, an electric field is induced inside in order to maintain an interior force-free condition. The charges accumulated at the surface in the process give rise to an external quadrupolar electric field whose component at the stellar surface (orders of magnitude stronger than gravitational forces for typical pulsar parameters) extracts charged particles from the surface. Strong electric fields cause efficient particle acceleration. The particles emit curvature radiation, and the radiation then decays into electron--positron pairs in strong magnetic fields near the pulsar surface \citep{Sturrock1971,rs75}. This phenomenon is referred to as ``pair production discharge''. Thus, the in-vacuo condition cannot be maintained and the pulsar magnetosphere gets filled with dense plasma. In the magnetosphere the electron-positron plasma co-rotates with the pulsar. This implies an outer bound beyond which the velocity required for co-rotation would exceed that of light. This imaginary cylindrical boundary is known as \textit{light-cylinder}. The light cylinder classifies the magnetic field lines into \textit{closed} (that close within the cylinder) and \textit{open} field lines. The \textit{last open field lines} define the edges of the \textit{polar caps} centred at the magnetic poles of the star. 

Despite forming an elegant picture, early models did not self-consistently predict the shape of the global magnetosphere at all scales, from near the pulsar surface to the light cylinder and beyond. Obtaining a closed solution proved to be an extremely non-trivial problem using analytical theory, and thus, it awaited advanced numerical simulations. The initial steps to solution, in the force-free electrodynamics (FFE) limit, were presented by \cite{Contopoulos1999}. Major advances have since been achieved following the observational revolution in the gamma-ray band brought about by the Fermi telescope \citep[e.g.][]{Smith2023}, which triggered an avalanche of simulation studies \citep[e.g.][]{Cerutti2024} aimed at understanding, simultaneously, the structure of magnetospheres and location of particle acceleration regions. Today, advances in modern computing have enabled full kinetic simulations and have hence opened up the opportunity to make direct investigation of pulsar magnetospheres at various scales. Modern simulation efforts focus on both local and global magnetosphere modelling and can include physics of particle acceleration, photon emission and pair production. Local targeted simulations are useful for understanding small-scale details of particle acceleration and emission physics, while global simulations aim at large-scale magnetospheric geometry and identification of particle acceleration regions. 

A goal of modern simulations and observations is to explain the diverse radio polarization features of pulsars as detailed in Sections \ref{sec:constrainalpha} and \ref{sec:morphpolspec}. A first simulation approach to this was attempted by \cite{Benacek2025}, who simulated the polarization properties produced by pair cascades in the polar cap. The results replicated many features of young pulsar polarization, but more complex polarization features were not replicated: this may be because they are the result of subsequent propagation effects in the magnetosphere not included in the model, as discussed in Section \ref{sec:morphpolspec}.

The key successes of simulation work in recent years are as follows. First, it has been found that pair cascades are time-dependent \citep{Timokhin2013} and can directly source polar radio emission \citep{Philippov2020, Cruz2021, Bransgrove2023, Benacek2024}. Second, high-energy emission can be produced in or close to the current sheet \citep{Cerutti2016,Kalapotharakos2018,Philippov2018} and so are (likely) the giant radio pulses \citep{Lyubarsky2019,Philippov2019}. Although many questions remain about connecting theory to observations, these represent dramatic advances in our understanding of the extreme physics of pulsar magnetospheres. 

The main limitations yet to be addressed are as follows. Simulations to date have primarily targeted the young pulsar population, meaning that investigations into the differences seen for milli-second pulsars, old pulsars and magnetars are still required. The problem of surface X-ray heating, particularly for millisecond pulsars, still needs to be addressed \citep[see details in section 4.5.1 of][]{Philippov2022}. Finally, mechanisms driving considerable observed time-variability of pulsar radio emission, as described in Section \ref{sec:timevar}, have yet to be uncovered by simulations. Continued progress in both observational and simulation-based investigation of pulsar physics is expected to significantly enhance our understanding of these questions in the coming years.

\subsection{Magnetosphere environment revealed by observations of the double pulsar}

In Section \ref{sec:constrainalpha}, we discussed how observations of the double pulsar system enabled precise constraints to be placed on its evolving system geometry. In addition, changes in the polarization state of pulsar A throughout the eclipses also provide a novel means of probing the extreme plasma environment surrounding pulsar B.
Observations of a change in linear-polarization position angle by \cite{ymb+2012} was interpreted as either the result of Faraday rotation from mildly-relativistic electrons in the magnetotail of pulsar B, or the preferential absorption of radio waves perpendicular to the local magnetic field direction at eclipses phase where emission from pulsar A grazes the closed-field region of pulsar B.
Recent observations with MeerKAT confirmed the latter effect, while also revealing enormous amounts of birefringence-induced circular polarization being generated at these grazing eclipse phases (see the purple curve in Figure~\ref{fig:double_psr_eclipse}), where the circular handedness is directly linked to the average magnetic field direction of pulsar B along the line of sight \citep{lkj+24}.  At present, these studies are limited only by telescope sensitivity and the fidelity of the magnetospheric models. The stunning sensitivity of the SKA will therefore enable the eclipses to be studied in unprecedented detail than is possible with any current facility. Being approximately three to four times more sensitive than MeerKAT at similar observing frequencies, the eclipse light curve presented in Figure~\ref{fig:double_psr_eclipse}, generated from averaging together four separate MeerKAT observations, would be equivalent to a single SKA-Mid observation in either the AA$^{\star}$/AA4 configurations. Measurements of the varying eclipse depth in broadband observations by both SKA-Low and SKA-Mid will yield better constraints on the radial distribution of charged particles within the magnetosphere and the transition frequency at which the plasma becomes transparent to radio waves \citep{bkm+2012}. These observations will in turn inform numerical simulations of neutron star magnetospheres, where comparisons between simulated and observed polarized radiative transfer within the eclipses can better constrain the properties of the confined plasma, such as pair multiplicity, magnetic field strength and the presence of heavy ions.

\subsection{Magnetospheric origin of subpulse drifting}

As discussed in Section \ref{sec:timevar}, one of the most striking observational features in pulsar radio emission is the existence of highly regular patterns of subpulse behaviour. For example, the ever-changing and subtle subpulse patterns revealed by the Thousand-Pulsar-Array (TPA) single pulse modulation survey \citep{Song2023} have yielded major clues to understanding the interaction between a pulsar's magnetosphere and its polar cap region. Observational studies have revealed a myriad of complex phenomena: subpulses which apparently drift both faster and slower than the pulsar rotation, frequently change slope and even reverse direction, always between pre-determined states \citep{Weltevrede2006, Weltevrede2007}. The emission can also suddenly cease (``nulling") and restart, often in a quasi-periodic fashion, or switch to a dramatic disordered mode then back to orderly drift.

Such patterns could be caused by spatial or temporal variations in the magnetosphere, and it may be argued that these effects result from subtle changes in the electric potential within the polar cap. However, it is known from those pulsars within the TPA survey which are observable at both poles (Song et al 2025, in preparation) that patterns at one pole can be communicated to the other – suggesting a strong interaction between polar and magnetospheric states \citep{Wright2022}. Early models assumed that the magnetosphere was in a permanent steady state and the polar cap region consisted of a single simple carousel of ``sparks'' \citep{rs75}, but up to now, even higher-dimensional simulations \citep[e.g.][]{2024ApJ...974L..32C} do not form or show the spatial emission regions (``sparks'') that have been qualitatively invoked to explain the subpulse formation. For the required plasma drift, however, the connection between simulations and observations has been made \citep[cf.][]{2012ApJ...752..155V}. Detailed study and modelling of TPA data is already underway (Hsu et al, submitted 2025) to develop an understanding of these phenomena, at least in quasi-geometric terms, which will ultimately lead to physical insight when compared with the predictions from theory and simulations. The sensitivity of the SKA telescopes, and hence the ability to overcome signal-to-noise-ratio limitations, will be essential to unambiguously address what physical parameters define the population of pulsars that show clear drifting, whilst the large number of new pulsars found and detailed with SKA will allow for both statistical and individual studies. 

\subsection{Pulsar wind nebulae}

A further measurement of note which is useful for comparison with modern simulations is that of the multiplicity in the neutron star magnetosphere. It is possible using pulsar wind nebulae (PWNe) to measure the total number of electrons and positrons in the PWN, which -- coupled with estimates of the pulsar's braking index and true age (which can also come from modelling the properties of the PWN) -- can lead to an estimate of its time-averaged multiplicity (\citealt{Spencer2025} and see \citealt{gnp+2025} from this special issue).

\subsection{Summary}
Modern simulations constitute a significant advance in our understanding of the underlying physics driving the generation of radio and multi-wavelength emission from pulsars. They have demonstrated that it is possible to generate radio emission and to have a self-consistent physical picture of a magnetosphere, however they cannot yet replicate the full complexity of observations, particularly the complex time-variability observed on all timescales. Furthermore, a simulation of a pulsar magnetosphere is, by definition, replicating a single example of a magnetosphere. Therefore we must be driven by observations to understand how physical parameters of pulsar magnetospheres (such as the magnetic field geometry) evolve across the population. This is the subject of the next section and the last of our five key open topics in pulsar magnetosphere physics.

\section{How do pulsars evolve across the population and over their lifetimes?}

\begin{figure}
    \centering
    \includegraphics[width=\columnwidth]{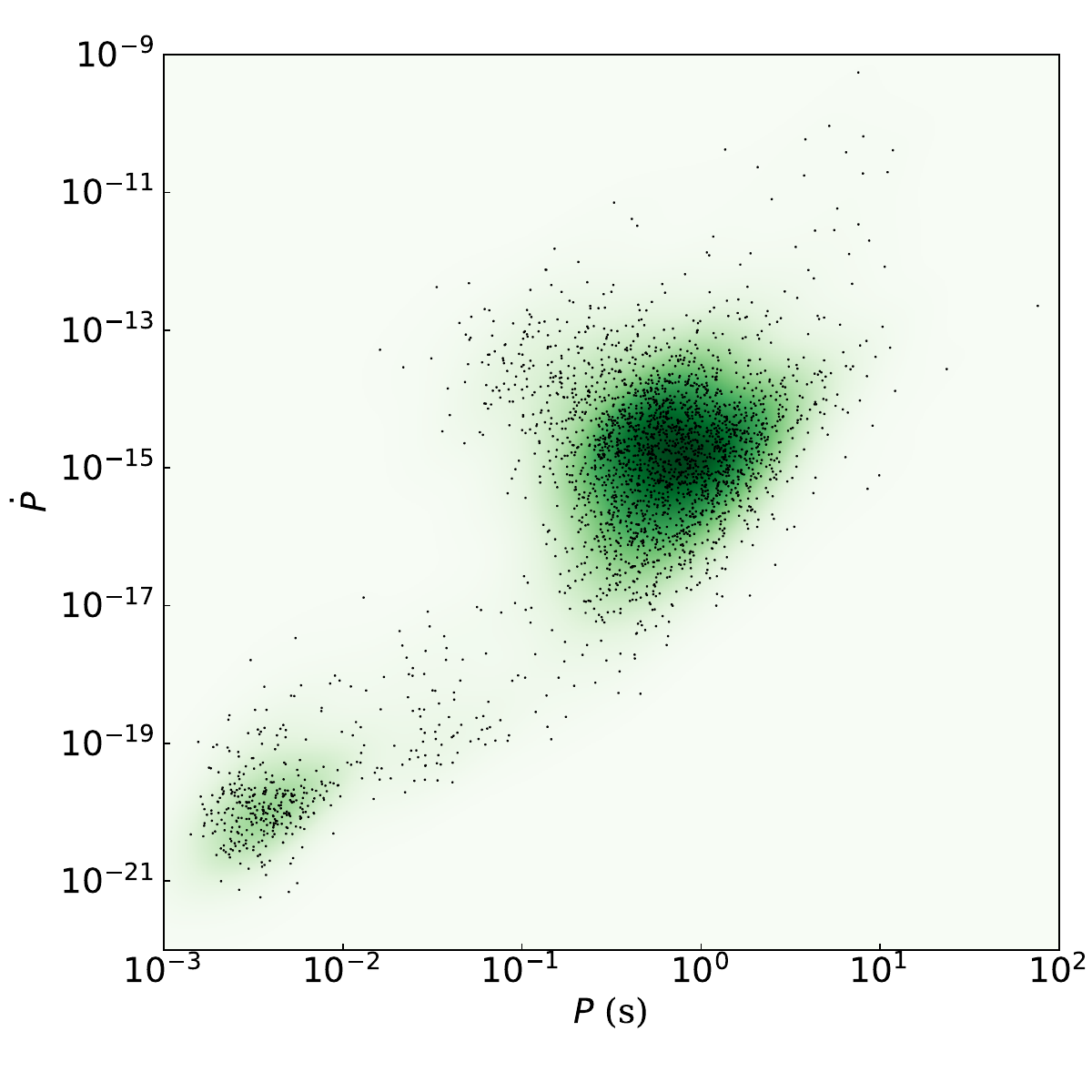}
    \caption{This figure shows all of the pulsars discovered to date which have both a measured period $P$ and period derivative $\dot{P}$, as a function of those two variables. The pulsars are shown with small black dots, and a Gaussian kernel density estimation of the whole distribution is shown in green. A total of 4,343 pulsars have been discovered, of which 2,816 have a measurement of $\dot{P}$. Data taken from the pulsar catalogue {\sc psrcat} \citep{Manchester2005} on 6th July 2025. }
    \label{fig:PPdot}
\end{figure}

Spin evolution presents a concise yet comprehensive picture of the life-cycle of pulsars. Being highly magnetized, rapidly rotating objects with very large moments of inertia, pulsars undergo spin-down due to loss of rotational kinetic energy through magnetic dipole radiation, winds and a small amount of radio emission \citep{lk12}. The spin period ($P$) and the spin-down rate ($\dot{P}$) are commonly used to provide estimates of the characteristic age ($\tau \propto P/\dot{P} $), surface magnetic field strength ($B \propto \sqrt{P\dot{P}}$) and spin-down luminosity ($\dot{E} \propto \dot{P}/P^3$) of pulsars. It should be noted that these estimates rely on the following assumptions in addition to the underlying model of a dipolar magnetic field: that the birth period of the pulsar should be much smaller than the current period, that there is no evolution over time of the magnetic field or beam geometry, that the surface magnetic field is well described by the spinning dipole model used to derive it, and that the spin-down is not excessively impacted by additional factors, such as precession or glitches. However, it has demonstrated for many pulsars that such assumptions do not hold over the timescales of measurement: whereas the simple model of magnetic dipole spin-down predicts a braking index of $n = 3$, published braking index measurements range from $n = -70$ \citep{Parthasarathy2020} to $n = 23,300$ \citep{Lower2023a}, and hundreds of pulsars have been found to show variable spin-down rates, frequently correlated with changes in profile shape as discussed in Section \ref{sec:timevar}. Similarly, pulsar characteristic ages are found to deviate from kinematic age measurements by up to a factor of four \citep{Suzuki2021}. 

Thus, the $P-\dot{P}$ diagram, as seen in Fig. \ref{fig:PPdot}, provides an elegant tool for demonstrating the evolution and classification of pulsars as a statistical whole, even if it cannot be used to precisely determine the properties of an individual source. The diagram broadly distinguishes between the two islands of points representing canonical pulsars ($P \sim 0.5$--$1$ s, $\dot{P} \sim 10^{-15}$ s~s$^{-1}$) in the upper right corner, and millisecond pulsars ($P \sim 3$ ms, $\dot{P} \sim 10^{-20}$ s~s$^{-1}$) in the lower left corner. Further attempts to classify pulsars into classes within the $P$--$\dot{P}$ diagram have been made in recent years using principal component analysis \citep{Garcia2022,Garcia2024} and unsupervised machine-learning \citep{Ay2020}. The spin-down luminosities are expected to be the highest for young pulsars with short rotational periods and large spin-down rates. Under this assumption, we would expect a canonical pulsar to be born in the upper left corner of the $P$--$\dot{P}$ diagram, lose its rotational kinetic energy rapidly, shift down to the island of points on the upper right side, and gradually spin down further beyond its characteristic age to cease any detectable dipole radiation, thus finally settling past the `death line' at the lower right corner of the plot which is also known as the `graveyard' zone. In recent years, long-period pulsars have been discovered which lie well beyond the death line as defined for a dipolar magnetic field geometry \citep[e.g.][]{Caleb2022}, as well as long-period radio transient sources discussed further in Section \ref{sec:lpts}, which raise challenges to the concept of the `death line' as a hard cut-off for radio emission. 

After this main evolution phase of the pulsar lifetime, it is thought that pulsars can transition to millisecond pulsars through accretion of matter and angular momentum from a companion star in a binary system. It is observed that, whereas ordinary pulsars are predominantly isolated, millisecond pulsars are mostly accompanied by companions. Millisecond pulsars with low-mass companions like white dwarfs are characterized by smaller spin periods and eccentricities, compared to those in double neutron star systems. Interestingly, it has been shown that, despite the considerable differences between the slow and millisecond pulsar populations in terms of spin properties, inferred ages, and life histories, many other observational properties are shared across the two populations, supporting common emission physics across the populations. This includes the same flux density spectra, the same dependencies of pulse width on period and of linear polarization fraction on spin-down energy, the same fraction of PA curves being described by the rotating vector model, and comparable inferred intrinsic luminosities \citep{Karastergiou2024}.

Modern surveys of the known pulsar population are on a sufficiently large scale to be able to extract correlations between average pulse profile parameters and properties derived from $P$ and $\dot{P}$. The TPA census was used to demonstrate new and update existing correlations between spin-down energy and the following parameters: radio pseudo-luminosity, flux density spectral index, linear polarization fraction and pulse width \citep{Posselt2023}. It also showed that correlations with characteristic age and inferred surface magnetic field were either weak or nonexistent. \cite{Oswald2023a} further investigated the average polarization properties of the pulsar population, demonstrating that polarization fraction, circular polarization contribution, profile complexity, and the frequency dependence of these properties, are all correlated with spin-down energy across the population. Studies suggest drifting subpulses are more likely to be found in older and less energetic pulsars \citep[e.g.][]{Weltevrede2006, Weltevrede2007, Basu2016, Basu2019, Song2023}. In particular, \citet{Basu2016, Basu2019} found that the modulation periods $P_3$ of the regulated drift patterns were anticorrelated with the spin-down luminosity $\dot{E}$ for a sample of around 100 pulsars, as low $\dot{E}$ pulsars often had large $P_3$s. With a much larger sample from the TPA of about 400 drifting pulsars, \citet{Song2023} suggested a V-shaped evolution of $P_3$ with characteristic age, where large $P_3$s were seen in the old and young population, and a minimum of $P_3\simeq2$ appeared around $10^{7.5}$ yr. Such evolution can be reproduced by a monotonic evolution of intrinsic $P_3$ due to the effect of aliasing, which naturally explains that younger pulsars have more erratic emission and aliased intrinsic $P_3<2$. As pulsars become older, $P_3$s become larger and the drift patterns become more organized: key evidence that magnetospheric processes evolve over the pulsar lifetime in a predictable way. 

Comparing observational trends from pulsar population surveys with predictions of spin evolution from the underlying physics is useful for understanding how both pulsar spin-down and magnetosphere structure may evolve over a pulsar's lifetime. Modern machine-learning methods and simulation-based inference are increasingly employed to perform pulsar population synthesis, to predict the evolution of pulsar physical properties that would lead to the $P-\dot{P}$ diagram as observed today \citep{Graber2024}, and to extend this to predict the numbers and properties of the new pulsars that the SKA telescopes will discover \citep[see][from this special issue]{Keane2025_SKA_Census}.

\subsection{Radio-loud magnetars}

Towards the upper-right of the $P$-$\dot{P}$ diagram reside the magnetars, slowly spinning ($P = 2$--$12$\,s) neutron stars that are thought to host the strongest magnetic fields in the Universe \citep{vb+2017}.
They are prolific X-rays and gamma-ray emitters, that are believed to be powered by the unwinding of and decay of their ultra-strong internal magnetic fields, as opposed to the spin-down energy that powers classical pulsars \citep{dc+1992}. 
This process induces enormous amounts of strain in the crust of these objects, leading to extremely energetic outbursts following crustal-fracture induced starquakes and magnetic reconnection events \citep{td+1995, l+2002}, detected by all-sky monitors as short hard X-ray/soft-gamma-ray flares. 
Initial surveys of magnetars using radio telescopes failed to detect pulsar-like radio emission from them \citep{lx+2000}, which led to speculation that pair-production in their magnetospheres is suppressed by their magnetic field strengths exceeding the quantum critical limit of $4.41 \times 10^{14}$\,G \citep{zh+2000}.
This radio-silent picture of magnetars changed with the detection of radio pulses from XTE~J1810$-$197 following its 2003 outburst \citep{crh+2006}.

Only five additional `radio-loud' magnetars have been discovered to date \citep{crh+2006, lbb+2010, efk+2013, ccc+2020, CHIME+2020}, plus SGR 1935+2154, for which transient radio pulses were detected over a brief period following its X-ray outburst in 2020 \citep{wlj+2024}. A seventh candidate radio-loud magnetar could be identified in pulsar J1119$-$6127, which is a radio pulsar that showed magnetar-like alterations in its radio emission following a glitch \citep{Weltevrede2011}. Whilst the radio pulses from magnetars share some similar behaviours with rotation-powered pulsars, such as orthogonal polarisation modes \citep{ksj+2007}, emission-state switching \citep{hgr+2008, ljs+2021}, and quasi-periodic microstructure \citep{Kramer2024}, they also exhibit several unique characteristics.
Their single pulses are comprised of many millisecond-duration sub-pulses that have drawn comparisons to fast radio bursts \citep{pmp+2018, dlb+2019, mjs+2019}, while their average profiles are subject to extreme amounts of spectral, temporal and polarimetric variability that manifest as the emergence or disappearance of profile components (e.g. \citealt{ksj+2007, scs+2017, ccc+2020}) and changes in spectral index \citep{ljs+2021, mscj+2022}. 
Propagation effects imprinted as frequency-dependent conversion between linear and circular polarisation offers a novel means of probing the plasma dynamics of their magnetospheres \citep{ksj+2007, ljl+2024}.
Each of these behaviors arise from the dynamic nature of their magnetic fields following an outburst, and are often associated with large variations in spin-down rate that are caused by the magnetic fields becoming increasingly twisted and then slowly unwinding as the magnetar relaxes back towards quiescence \citep{b+2009, scs+2017, lld+2019}.
Quasi-periodic changes in the linear polarisation position angle of XTE~J1810$-$197 following its 2018 outburst may have been caused by the magnetar undergoing damped free-precession, and it is confirmed that this would provide additional insights into the interface between the neutron star crust and magnetic field \citep{dwg+2024}.
Broadband studies of these objects, both known and yet-to-be identified, with the SKA-Mid telescope in sub-array mode covering the 0.3-15\,GHz range of Bands 1, 2 and 5, will provide unique insights into the impulse response of neutron star magnetospheres following explosive events. 

The enormous volume of pulsars that will be discovered through the SKA-Mid pulsar surveys should include several new radio-loud magnetars in X-ray quiescence, similar to PSR~J1622$-$4950 which was found during a blind radio pulsar survey at Murriyang \citep{lbb+2010}.
Magnetars have characteristically flat radio spectra and are among the brightest radio pulsars at frequencies $>$10-100 GHz \citep{tbb+2022} with low Galactic scale heights. This makes high-frequency SKA-Mid surveys of the Galactic plane particularly conducive to finding them.
The transient nature of their radio emission means that most of these objects are identified following outbursts, motivating repeat surveys and rapid follow-up of high-energy detections.
Leveraging the unprecedented, broadband sensitivity of the SKA to target known, seemingly `radio-silent' magnetars, may uncover a population that exhibit comparatively weak radio emission, similar to that detected in FAST observations of SGR~1935$+$2154 \citep{zxz+2023, wlj+2024}.

\subsection{Links to Fast Radio Bursts} \label{sec:frb}

Increasing observational evidence points towards neutron stars being a potential originator of Fast Radio Bursts (FRBs). In particular, the observation of extremely bright radio bursts from the Galactic magnetar SGR 1935+2154 cemented the idea of a magnetar origin of FRBs \citep{CHIME+2020}. High time-resolution studies of Fast Radio Bursts (FRBs), at frequencies high enough to avoid most scattering and dispersion, find that these enigmatic bursts resemble the bursts seen in  young, energetic, highly magnetised neutron stars \citep{PastorMarazuela2025}. In particular, magnetars are among the most energetic neutron stars and interest in their properties has recently been heightened by them being a possible source of FRBs \citep{ww20}. Recently, \citet{Kramer2024} reported on detection of quasi-periodic sub-structure emission in most radio-loud magnetars. This feature has so far been seen in all classes of rotation-powered neutron stars, including canonical pulsars, millisecond pulsars and rotational radio transients \citep[e.g.][]{ccd68,mar15,Liu2022,dys+24}. The relation between rotational period of the pulsar and the quasi-periodicity can be modelled by a simple power law that is shown to expand six orders of magnitude \citep{Kramer2024}. The measured quasi-periodicity of the sub-structure in magnetars also nicely obeys this power-law relation. Additionally, quasi-periodic sub-structure was reported for a small number of bursts from FRBs \citep[e.g.,][]{abb+22,pvb+23}. Therefore, by assuming a neutron-star origin, the power-law relation may be used to infer the potential period in an FRB. However, investigation along this line is greatly limited by the number of case studies at the moment. The SKA will find greater numbers of both FRBs and young pulsars, and allow for further comparison in their time, frequency, polarisation, and localisation behavior. More firmly linking the two would help solve one of the main open questions in present-day astronomy: the nature of FRBs.

\subsection{Long-period radio transients} \label{sec:lpts}

Recent years have seen the discovery of both pulsars with very long rotational periods, such as the 76-s pulsar \citep{Caleb2022}, and of long period radio transients \citep[LPTs, e.g.][]{Hurley-Walker2022}. Long-period pulsars raise a challenge to simplistic pulsar evolution models, since they occupy a region of $P-\dot{P}$ space well beyond the ``death line'' described above. LPTs have even longer periods (up to 6.45 hours discovered so far) \citep{Lee2025} and it is as yet unclear whether these also originate from neutron stars, or from an alternative source or physical mechanism, such as from White Dwarfs \citep{Rea2024}. Some LPTs with detected orbital periodicities have indeed been proved to originate from White Dwarf binaries \citep[e.g.][]{DeRuiter2025} but others show observational similarities to magnetars and FRBs \citep[e.g.][]{Men2025}. In the magnetar scenario, a possible explanation may be that crustal movement in magnetars leads to local magnetospheric twisting which in turn results in coherent radio emission \citep{Cooper2024}. It is clear that more detailed study of radio pulsars, radio-emitting magnetars and of LPTs is required to fully understand the similarities and differences between their observational features and hence probe the mechanism driving radio emission in the magnetosphere in each case. New observations with more advanced instrumentation are opening up the parameter space of radio transient physics and propelling advancements in understanding pulsar magnetospheres. New discoveries with the SKA telescopes will advance this still further. 

\subsection{Summary}
Studying the statistical trends of the observed pulsar population reveals information about how neutron stars evolve over their lifetimes. It also provides a robust comparison to new discoveries, of pulsars and other radio transients including FRBs and LPTs, which push the boundaries of our understanding of magnetosphere physics. Large-scale pulsar monitoring with the SKA presents an exciting opportunity to probe the population physics of neutron stars to a greater depth than ever previously possible.

\section{Opportunities with the SKA telescopes}

In this paper we have presented the recent advances in answering the five key open questions in pulsar magnetosphere science, and the ways in which future observations with the SKA telescopes will enable progress in this field. The key topics to be addressed are understanding the magnetic field geometry, emission spectrum and time-variability of radio pulsars, and placing this information into the context of both the global magnetosphere physics, and the pulsar and radio transient population as a collective whole. 

The SKA telescopes will be ideal in providing the essential radio telescope requirements to advance our understanding of these areas. The high instantaneous sensitivity and band-widths, high fidelity polarization calibration, and combination of high and low frequency receivers across SKA-Low and SKA-Mid will ensure outstanding quality radio observations of pulsars. The telescopes' considerable flexibility as instruments, particularly their capacity for sub-arraying, will enable high-quality monitoring of the pulsar population to obtain a collective view of pulsar behaviour. The large field of view of SKA-low, coupled with a automated detection pipeline, will be helpful in triggering dense cadence monitoring of rare events to follow post-event radiative changes. Combining this with regular cadence monitoring, simultaneous observing with facilities at other wavelengths, and the potential for quick reactivity to interesting events, would assure optimal advances in pulsar magnetosphere science in the coming years.

In the remainder of this section we address each of the observational requirements of pulsar magnetosphere physics in detail and explain the unique observational possibilities that the SKA telescopes will provide. We then present a summary of the ideal observational approach to maximise the science potential from the AA* configuration.

\subsection{Unique possibilities}

New discoveries widen the available exploratory phase space for new physics. This has been made clear in recent years not only from the vast increase in the number of pulsars discovered, with 4,343\footnote{Source: ATNF pulsar catalogue \citep{Manchester2005}, queried on 6th July 2025.} now known compared to 2,519 ten years ago, but also from the expanded discovery of FRBs and LPTs, and the observational features connecting these populations to neutron stars. Discoveries of individual unusual sources, such as the double pulsar, also have the potential to transform our understanding of pulsar magnetosphere physics. This in turn drives simultaneous advances in both observations and theory, ultimately enabling us to connect the two. \textbf{Pulsar searches with the SKA telescopes will dramatically expand the observed population of radio pulsars and will discover new and unexpected behaviour, changing our perspective of the pulsar magnetosphere.}

Large-scale surveys at a wide range of radio frequencies, such as the Thousand-Pulsar-Array survey, the Parkes Young Pulsar Array, and monitoring projects with the LOFAR and GMRT telescopes, enable a population-wide view of pulsars. Such a view allows the extraction of underlying trends, revealing how neutron stars evolve over their lifetimes. \textbf{Pulsar monitoring with the SKA telescopes will benefit from broad-band observations with both SKA-Mid and SKA-Low, as well as subarraying capacity to simultaneously observe multiple pulsars. This will make it possible to build up a connected picture of the whole population and the physics of its time-variability.}

Highly sensitive, broad-band observations enable probes of radio emission on an unprecedented level which open new questions to be addressed. Recent examples have included new discoveries of subtle profile variability on long timescales, detailed single-pulse studies at high time resolution, and significant advances in understanding pulsar magnetosphere composition through targeted observations of the double pulsar. Increased sensitivity also assures the systematic detection of weak pulse components with fluxes below a few per cent of the peak flux \citep{Xu2025}. This advances our ability to probe the magnetosphere across a much wider range of longitudes, providing a more complete picture of how radiation is generated within the magnetosphere. \textbf{The sensitivity of the SKA telescopes will make them the ideal instruments for carefully targeted follow-up of the most promising and unusual radio pulsars, to probe their magnetospheric physics in depth.}

As a sensitive interferometer, the Square Kilometre Array (SKA) will uniquely provide the opportunity to resolve pulsar scattering screen structures. The interference between speckles on these screens achieves astrometric accuracy of 100 picoarcseconds \citep{Pen2014}. With its enhanced sensitivity, the SKA telescopes will increase the number of pulsars whose scattering screens can be resolved. \textbf{Using the SKA telescopes as highly sensitive interferometers will significantly improve ongoing efforts to resolve pulsar emission regions \citep{Lin2023}, paving the way for full image reconstruction of these regions.}

\subsection{Observing modes with AA*}

Pulsar magnetosphere physics will be best advanced through the following observations with the SKA: 
\begin{enumerate}
    \item A large-scale survey of the pulsar population to understand broad statistical trends, which exploits its broad-band observational capacity with SKA-Low and SKA-Mid;
    \item Ongoing monitoring and timing of a large sample of pulsars to probe time-variability on all time-scales;
    \item In-depth studies of individual sources of particular interest, such as the double pulsar;
    \item Targeted follow-up of new pulsar discoveries, FRBs, and LPTs, to constrain their properties; and
    \item Simultaneous observing with facilities at other wavelengths, particularly X-rays and gamma-rays. 
\end{enumerate}

Optimal observations would be recorded in a combination of both search- and fold-mode, and in full Stokes to ensure polarization studies could be completed. Collecting high signal-to-noise ratio observations of single pulses across a wide frequency band (encompassing both SKA-Low and SKA-Mid) will be very important to simultaneously understand the frequency- and time-variability of pulsar radio emission: key examples would include collecting single pulses from millisecond pulsars in support of advancing pulsar timing array accuracy, and mapping polarization evolution with frequency. 

A pulsar survey with the SKA could take inspiration from the Thousand-Pulsar-Array (TPA) survey on the MeerKAT telescope, which has collected high quality observations of over a thousand pulsars, and monitored 597 of those sources on a monthly basis since 2019. The TPA survey made use of sub-array capabilities and carefully defined the observing strategy to simultaneously maximise observational fidelity and observing efficiency \citep{Song2021}. The observing goals for pulsar magnetosphere physics overlap strongly with those for studying galactic plasmas with pulsars \citep[see][from this special issue]{Tiburzi2025_SKA_Plasma}, and for surveying the neutron star population \citep[see][from this special issue]{Levin2025_SKA_NSpop}, as well as the key goal of following up new discoveries made through SKA pulsar searches \citep[see][from this special issue]{Keane2025_SKA_Census}. These in turn will help better characterize the noise sources in a pulsar timing experiment, thereby significantly improving the sensitivity of the SKA Pulsar Timing Array \citep[see][from this special issue]{Shannon2025_SKA_SKAPTA}.

The AA* array configurations are to deploy 307 of the 512 SKA-Low stations, and 80 of the 133 SKA-Mid dishes which, combined with the 64 existing MeerKAT dishes, makes 144 in total for SKA-Mid at this stage. The missing stations will predominantly impact intermediate baselines for Low and maximum baselines for Mid, but since the pulsar observations proposed do not involve interferometric imaging (except in the specific use case of resolving pulsar scattering screens), the only constraint is on overall sensitivity of the array. Ultimately this will mean that full deployment of AA4 will expand the number of pulsars for which high-quality in-depth observations can be taken, by improving the signal-to-noise ratio of individual single pulses in weaker and more distant pulsars.

The final key components of SKA observing that will uniquely advance studies of pulsar magnetospheres with this instrument are as follows. High-frequency observations at 5--10~GHz will transform our understanding of the frequency-dependence of pulsar radio emission, and SKA-Mid will be uniquely sensitive at this frequency range, combating the loss of flux density at higher frequencies due to pulsars' steep flux spectral indices. The SKA telescopes' positioning in the Southern Hemisphere opens up the galactic plane for observation and makes it the ideal instrument for high-quality observations of important targets in the Southern sky, such as the double pulsar. Finally, the world-leading collecting area of SKA-Low will transform our capacity to survey and monitor pulsars, advancing our understanding of the pulsar population as a cohesive whole.

\section{Conclusions}
In summary, this paper has detailed the considerable advances made in pulsar magnetosphere physics in the last decade, which have been achieved through a combination of detailed studies of individual sources, a survey-based approach to understand population trends, and the comparison of radio observations with both simulations and with observations at mm, UV, X-ray and gamma-ray wavelengths. The SKA telescopes will be uniquely suited both to monitoring a large number of pulsars with great sensitivity, and to be used for in-depth studies to address specific goals. This paper therefore lays out the importance of running a large-scale pulsar monitoring survey with the SKA telescopes, augmented by high time-resolution studies of specific sources, to achieve a full picture of pulsar magnetosphere physics in the coming years.

\section*{Acknowledgements}
LSO acknowledges support from the EPSRC Stephen Hawking Fellowship grant EP/Z534730/1. 
MEL is supported by an Australian Research Council (ARC) Discovery Early Career Research Award DE250100508.
XS and JVL acknowledge support from CORTEX (NWA.1160.18.316), under the research programme NWA-ORC, financed by the Dutch Research Council (NWO).
ALW acknowledges support from ERC Consolidator Grant No.~865768 AEONS. NR is supported by the European Research Council (ERC CoG No. 817661 and ERC PoC No. 101189496), and grants SGR2021-01269, ID2023-153099NA-I00, and CEX2020-001058-M.
JB acknowledges support from the German Science Foundation (DFG) Project No. BE 7886/2-1. 
AB acknowledges SERB (SB/SRS/2022-23/124/PS) for financial support. 
BCJ acknowledges the support from Raja Ramanna Chair fellowship of the Department of Atomic Energy, Government of India (RRC - Track I Grant 3/3401 Atomic Energy Research 00 004 Research and Development 27 02 31 1002//2/2023/RRC/R\&D-II/13886 and 1002/2/2023/RRC/R\&D-II/14369).

\bibliographystyle{aasjournal}

\bibliography{Article}


\end{document}